\documentclass[preprint,onecolumn,superscriptaddress]{revtex4}

\begin{document}

\title{Nebula-Relay Hypothesis: Primitive Life in Nebula and Origin of Life on Earth}
\author{Lei~Feng}
\email{fenglei@pmo.ac.cn}
\affiliation{Key Laboratory of Dark Matter and Space Astronomy, Purple Mountain Observatory, Chinese Academy of Sciences, Nanjing 210023}
\affiliation{Joint Center for Particle, Nuclear Physics and Cosmology,  Nanjing University -- Purple Mountain Observatory,  Nanjing  210093, China}

\begin{abstract}
A modified version of panspermia theory, named Nebula-Relay hypothesis or local panspermia, is introduced to explain the origin of life on Earth. Primitive life, acting as the seeds of life on Earth, originated at pre-solar epoch through physicochemical processes and then filled in the pre-solar nebula after the death of pre-solar star. Then the history of life on the Earth can be divided into three epochs: the formation of primitive life in the pre-solar epoch; pre-solar nebula epoch; the formation of solar system and the Earth age of life. The main prediction of our model is that primitive life existed in the pre-solar nebula (even in the current nebulas) and the celestial body formed therein (i.e. solar system).
\end{abstract}
\maketitle
\section{Introduction}

Generally speaking, there are several types of models to interpret the origin of life on Earth.
The two most persuasive and popular models are the abiogenesis~\cite{Oparin,Haldane} and panspermia theory~\cite{Panspermia}.

The modern version of abiogenesis is also known as chemical origin theory introduced by Oparin in the 1920s~\cite{Oparin},
and Haldane proposed a similar theory independently~\cite{Haldane} at almost the same time. In this theory, the organic compounds are naturally produced from inorganic matters through physicochemical processes and then reassembled into much more complex living creatures.
The famous Miller-Urey experiment~\cite{MillerUrey} and subsequent researches proved that several kinds of amino acids (the chemical constituents of the proteins) can be synthesized in the inorganic environment of the early Earth, which is a very positive evidence of the abiogenesis theory. Although the abiogenesis model has been widely recognized among scientists, many details of the entire process are far from clear.

The age of the Earth is about 4.54 billion years~\cite{earthage} and the liquid ocean was formed at about 4.325 billion years ago~\cite{oceantime}. There are solid evidences that life on the Earth existed at least 3.5 billion years~\cite{lifetime,lifetime2}.
Moreover, some researches suggested that the oldest life may have appeared as early as 4.28--3.77 billion years ago~\cite{lifetime3}.
Between 4.1--3.8 billion years ago, it is believed that Earth endured a series of powerful meteor shower.
The beginning of life on Earth occurred before or soon after this heavy bombardment period, i.e. there was only hundreds of millions of years to complete the formation of life on Earth. It is shorter than what is generally believed and might mean that the formation of life did not happened on Earth.

Panspermia theory was firstly suggested by Lord Kelvin in 1871.
In this theory, it is presumed that primitive lives, the seeds of Earth's
life, exist in the solar system, Milky Way and even the entire universe. And they propagated to the Earth through meteorites from other planets in solar system or the comets from other planet systems in the Milky Way.
The seeds of life must have the ability to survive in extremely harsh environment and might be one kind of extremophile~\cite{extremophile1,extremophile2}.

Even if this theory is real, it does not solve the problem of where and how life originated in the universe.
If the seeds of life came from any region of the Milky Way with different genetic coding system and if they can all arrived at the early Earth, the diversity of Earth's life should be much greater. But as we know, the creatures on the Earth share the same genetic-expression system and therefore have a common origin which usually named as ``the last universal common ancestor (LUCA)''. The first two hypotheses cannot be true at the same time. Of course, it can be attributed to the rarity and occasionality about the arrival of life-bearing systems from other planet systems. But the origin of life on Earth is completely an accident and such model is not convincing.

In this draft, we proposed a new model, named as Nebula-Relay hypothesis or local panspermia, which is actually a modified version of panspermia theory. It systematically solves the difficulties of aforementioned models and has clear predictions for further testes.

This paper is organized as follows: In Sec.2 we present our method. The main predictions of this model are presented in Sec.3 and the conclusions are summarized in Sec.4.

\section{Nebula-Relay Hypothesis}

In abiogenesis theory, the rapid emergence of life on Earth may indicate that the origin of life is not so complicated.
And in the panspermia theory, the seeds of life are somehow universal in the Milky way. In these two models, it can be expected that extraterrestrial civilizations are common in universe. However, the Fermi paradox tells us that such overestimated probability of extraterrestrial civilizations in the Milky Way contradicts with the lack of evidence. In ref.~\cite{Sandberg}, the authors found that the substantial probability that human is the only intelligent creature in the Milky Way is about 53\%--99.6\% and
39\%--85\% for the observable universe situation. In other words, the number of high-level intelligent creatures may be much less than the previous expectation.
Then the interpretation of Fermi paradox might be the rarity of high-level intelligent creatures.
This may imply that the occurrence of life is very complex and cost much longer time than the previous expectations.

In our model, the first and most important assumption is that the formation of seeds of Earth's life takes billions of years and it's much longer than the time between the birth of the Earth and life therein.
But where did this process happen? To explain the homology of Earth's life and avoid the contingency of panspermia theory, we should develop a more reasonable theory.

The sun is thought to be a third-generation star in the universe and formed from the pre-solar nebula, a molecular cloud, which is produced by the supernova explosion of sun's predecessor star~\cite{solar}. In this draft, we named the sun's predecessor epoch as the pre-solar epoch.
About the sun's predecessor star, we know very little about it.
Chemical elements composing of life, such as carbon, nitrogen and oxygen, are produced by the nuclear reaction process in the center of massive stars. So rocky planets and primitive lives can only be formed after the death of first-generation stars. Fortunately, as the second-generation star, the predecessor of sun can provide necessary materials for the origin of life.

The second assumption in our model is that primitive life occurred in the planet of sun's predecessor star and could live in the environment of molecular clouds.
Generally speaking, molecular clouds act as a relay station connecting the life in Solar System and its predecessor, so we call this model the Nebula-Relay hypothesis. Then the history of Earth's life can be divided into three epochs as follows.

\subsection{Epoch I: The formation of primitive life in the planet of pre-solar star}

In this epoch, primitive life was formed on one planet of the pre-solar star through complex physicochemical interactions.
It is extremely difficult to explore this process in detail now since we can't accurately know the environment at that time.
After formation, primitive lives could be dislodged into interstellar space by some special mechanisms. In ref.~\cite{Berera2017}, the authors pointed out that interplanetary space dust has enough kinetic energy to propel the microbial life of atmospheres into interstellar space. Some of them could then be blown far away from the pre-solar star though solar wind or other processes. This may be the reason how primitive life survive from the supernova explosion of pre-solar star. Moreover, if primitive life is wrapped up in some celestial bodies, such as comets and meteorites, its survival probability would be greatly enhanced.

Primitive life here doesn't have to refer to one specific species. It could also be a collection of creatures which have same origins and share some common characteristics.
Generally speaking, the simpler biological structure, the stronger environmental adaptability. Therefore we expected that primitive life have very simple structure.

We prefer that primitive life satisfy the RNA (Ribonucleic Acid) world hypothesis which was proposed by Alexander Rich~\cite{RNA1} and named by Walter Gilbert~\cite{RNA2}.
This is because the RNA creatures have much simpler structures and avoid the fragility of protein and DNA (DeoxyriboNucleic Acid).

We do not rule out the possibility that there were much more complex creatures in Epoch I. But such creatures do not adapt to the harsh environment of molecular cloud and have nothing to do with the life on Earth.

\subsection{Epoch II: Primitive life in the pre-solar nebula}

After the death of pre-solar star, primitive life-forms distributed in the produced pre-solar nebula, i.e. molecular cloud, and lived miraculously in such severe conditions. The reproduction of these primitive life in molecular cloud makes their density large enough when the Solar System was born.

Molecular cloud is a type of interstellar cloud and mainly composed of gases and dusts. Its typical temperature is about dozens of Kelvin. The average density is about $\rm 10^2-10^4$ molecules per cubic centimeter and hydrogen molecule is the most population constituent. In addition, many kinds of organic compounds were also found in molecular clouds, such as polycyclic aromatic hydrocarbons~\cite{PAHs}, fullerenes~\cite{fullerenes}, glycolaldehyde~\cite{glycolaldehyde} and so on.
Glycolaldehyde, a specific sugar molecule, plays an important role in the formation of RNA. These organics provide the necessary materials for the survival and reproduction of primitive life. It is reasonable to assume that there are similar compositions in pre-solar nebula.

In the low temperature environment of molecular cloud, primitive life maintains low biological activity and requires only a small amount of energy. A special biological mechanism is needed to utilize the energy in the molecular cloud.

Cosmic ray (CR) may be the possible energy sources, such as CR electrons and high energy photons.
Free radical might be one possible intermediate carrier of energy flow process in life.
If CR electrons are captured by the compounds in primitive life, active free radicals are formed naturally.
Another possibility is that free radicals are produced by the ionization processes of charged CR particles in primitive life. Further reactions of free radicals release its energy for primitive life. As pointed out in ref.~\cite{atp}, free radicals probably take part in the phosphorylation of adenosine diphosphate. The biochemical process of extremophile may provide us more useful information about such mechanism as they live in similar environment as primitive lives.
Moreover, it is conceivable to use organic macromolecules (similar to green fluorescent protein) to realize the energy conversion from high-energy electromagnetic radiation to the organisms in molecular cloud.

CR in molecular clouds contains two components which are primary component produced by CR source and secondary component produced by the collision of CR particles and molecular cloud. The collision process can also reduce the number of high-energy CR particles, which would protect primitive life from lethal radiation. The density of secondary CR particles ($q_{\rm sec}(p)$) at momentum $p$ can be calculated as follows:
\begin{equation}
  q_{\rm sec}(p)=\beta c n_{\rm H} \int {\rm d}p^\prime\frac{{\rm d}\sigma(p,p^\prime)}{{\rm d}p} n(p^\prime),
\label{sec}
\end{equation}
where $p^\prime$ is the momentum of primary CR nucleus. $n_{\rm H}$ and $n(p^\prime)$ are
the density of molecular clouds and CR particles, respectively. $v$ and $c$ are the velocity of CR particles and light, respectively. The factor $\beta (=v/c)$ presents the
consequence of random-walk process of CR in galactic magnetic fields and ${\rm d}\sigma(p, p^\prime)/{\rm d}p$ is the production cross section. The density of CR depends on the frequency of supernova explosions and it is hard to estimate the specific value at pre-solar epoch.

The energy density fixed by primitive life ($I^{\rm e,\gamma,...}$) from CR electrons, photons and other components is given by the following equation
\begin{equation}
I^{\rm e,\gamma,...}=\int {\rm d}p b^{\rm e,\gamma,...}(p) \left[q^{\rm e,\gamma,...}_{\rm pr}(p)+q^{\rm e,\gamma,...}_{\rm sec}(p)\right],
\label{flux}
\end{equation}
where $b^{\rm e,\gamma,...}$ is the energy conversion fraction of primitive life, $q^{\rm e,\gamma,...}_{\rm pr}(p)$ and $q^{\rm e,\gamma,...}_{\rm sec}(p)$ are the primary (pr) and secondary (sec) CR density at momentum $p$ for different CR components, respectively. The superscript e and $\gamma$ represent electrons and photons, respectively.
Furthermore, energy storage process in primitive life may also play an important role for its survival in nebula.
The transmission of CR electrons in the living body drives the formation of an electrochemical proton gradient and finally produces ATP (adenosine triphosphate) through molecular motors.

In short, we believe that it is possible for primitive life to survive in the molecular cloud. The fact that the ancestors of Earth's life lived in molecular clouds may be responsible for many magical phenomena in cryobiology and it may also relate to the indispensability of free radicals in current lives.

\subsection{Epoch III: The formation of Solar System and the beginning of life on Earth}

Primitive lives were wrapped into the celestial bodies of Solar System formed in pre-solar nebula. So they can be in all the suitable celestial bodies or even in the interplanetary  space because some of them are blown into the interstellar space when comets graze the sun.

The environment of the freshly formed Earth was particularly harsh and not suitable for survival. The volcanos
erupted very frequently and lava flowed on the ground. Hence the temperature was so high that there was no liquid water.
Primitive life might be brought to the Earth when (together with water by comets) or after the ocean was formed.
Anyway, parts of them arrived at Earth and enliven immediately acting as LUCA. Then LUCA began their evolutionary journey on Earth.
In unsuitable environment, some primitive lives could turn into fossils and they can be found in the current planets and their satellites, dwarf planets, comets and asteroids of Solar System.

If another star was formed in the pre-solar nebula, primitive lives and their descendants can also be found in its planetary systems.


\subsection{Differences between Nebula-Relay hypothesis and panspermia}

From the above discussions, we can see that our model is somehow one kind of panspermia theory. But it also has its own unique characteristics. For ordinary panspermia theory, life originates on other planet or other planetary system and transfers to Earth through some space transfer mechanism. But in Nebula-Relay hypothesis, life occurred in the local place at the pre-solar epoch and evolved for about several billions years in pre-solar nebula to the formation of the solar system. So we also named this model as local panspermia. In other words, our model clearly indicates the origin place of primitive life, which is not available in the panspermia theory.

For panspermia theory, the carrier of seeds of Earth's life is disposable and we can't find it now. Even if a similar vehicle carries new seeds to the solar system now, it has nothing to do with the origin of life on Earth.
However in the Nebula-Relay theory, all the objects in the solar system can be the carriers of seeds and we can still look for the traces of these primitive lives nowadays.

\section{Predictions and Enlightenments}


Searching for the vestiges of ever biological activity in the pre-solar nebula directly is in a sense impossible. But such primitive creatures or/and its offsprings are universal in the solar system. We can look for their descendants or fossils in objects formed in the pre-solar nebula, i.e. all the celestial bodies in the solar system, for second best. Several possible evidences or hints had already been reported.

In ref.~~\cite{meteorites2019}, the authors investigated three primitive meteorites which are believed to be more than 4.5 billion years old and formed in the early solar system. These ancient meteorites came from the protoplanetary disk which formed in pre-solar nebula. They found that there are riboses in these meteorites which is an essential component to build RNA. Thus we suspected that the riboses are the remnants of RNA creatures in pre-solar nebula.

The most ideal celestial body to test this model is Europa, Jupiter's satellite, because there are underground oceans under its thick ice sheet. There are same seeds on Europa as the early Earth and its thick ice sheet blocked the continuous input of seeds coming from other planets in solar system or other planetary system in Milky Way.
If similar DNA/RNA/protein life forms are found on Europa, it would be a potential evidence for our model. If not, it is a strong challenge for our model.

In our model, primitive life could survive in the molecular cloud environment. And perhaps, life also appeared in the predecessor stellar system of current molecular clouds.
Hence there may be primitive life in the present molecular clouds and it is a smoking gun for our model.

In ref.~\cite{biosignature}, the authors suggest that organosulfur compounds, particularly methanethiol (such as $\rm PH_3$, $\rm CH_3SH$ and so on), might be the biosignatures of extrasolar Earth-like planets. These biosignatures may also be used to search for primitive life in the nebula.

Lots of researches had been done about the spectral signatures of RNA molecules. For example, there is a very strong absorption peak in the ultraviolet band and the spectra of RNA nucleobases are in the terahertz band~\cite{terahertz,terahertz2,rna-thz,rna-thz2}. In refs.~\cite{terahertz,terahertz2}, the authors pointed out that the terahertz circular dichroism spectroscopy can be applied to investigate the metabolic and genetic machinery of extraterrestrial life. Searching nucleotides in molecular clouds with these methods could verify the conclusion indirectly that life can exist there.

\section{Summary and Discussions}

In this draft, we introduced a new hypothesis, Nebula-Relay hypothesis or local panspermia, to interpret the origin of life on Earth.
In our model, primitive life was formed at the pre-solar epoch and spread to the pre-solar nebula. As the Solar System was born, these primitive lives were coerced into the produced celestial bodies. The journey of Earth's life is a great miracle full of many uncertainties.

Our model has some advantages comparing with the previous ones. It avoids the suspected difficulty of the beginning time of Earth's life and diversity problem of panspermia theory. It should be pointed out that there are clear theoretical expectations to be identified in our model which is discussed in detail in the third section.

This paper is somehow like a framework or roadmap about the origin of Earth's life and too many details are still unclear. Anyway, it is valuable to point out such theoretical possibility.

\acknowledgments
We thank Dr. Yuan-Yuan Chen, Shi-Yong Liao and Zi-Qing Xia for helpful discussions and suggestions. This work is supported by the National Key Research and Development Program of China (Grant
No. 2016YFA0400200); the National Natural Science Foundation of China (Grants No. 11773075) and the Youth Innovation Promotion Association of Chinese
Academy of Sciences (Grant No. 2016288).




\renewcommand{\refname}{References}

\end{document}